| | |
|---|---|
| Title | Reactivity of water vapor in an atmospheric argon flowing post-discharge plasma torch |
| Authors | S Collette, T Dufour and F Reniers |
| Affiliations | Faculté des Sciences, Service de Chimie Analytique et de chimie des Interfaces, Université Libre de Bruxelles, CP-255, Bld du Triomphe, B-1050 Bruxelles, Belgium |
| Ref. | Plasma Sources Sci. Technol. 25 (2016) 025014 (11pp) |
| DOI | 10.1088/0963-0252/25/2/025014 |
| Abstract | The reactivity of water vapor introduced in the flowing post-discharge of an RF atmospheric plasma torch is investigated through electrical characterization, optical emission spectroscopy and mass spectrometry measurements. Due to the technical features of the plasma torch, the post-discharge can be considered as divided into two regions: an inner region (inside the plasma torch device) where the water vapor is injected and an outer region which directly interacts with the ambient air. The main reactions induced by the injection of water vapor are identified as well as those indicative of the influence of the ambient air. Plausible pathways allowing the production of H, OH, O radicals and $H_2O_2$ are discussed as well as reactions potentially responsible for inhomogeneities and for a low DC current measured in the flowing post-discharge.

# Introduction

At first inspection, traces of water vapor in a plasma source may be considered as problematic since they can induce undesired chemical reactions [1, 2] and often instabilities [3]. However, for specific applications, water vapor can be deliberately mixed with the plasmagen gas either to achieve a milder treatment or to generate radicals of interest. This approach has already been investigated in various domains such as surface treatments [4], biocompatibility [5] or plasma medicine [6].

In surface treatments, the oxidative functionalization is usually achieved with $O_2$ [7–11]. This reactive gas can, however, be replaced by water vapor and mixed with a carrier gas to induce a milder treatment. For instance, the exposure of graphene to a pure oxygen plasma at 0.3 mbar leads to an over-etching of the top sheets even on short exposure times (few seconds) while the addition of water-vapor to the plasma allows a milder and more homogeneous treatment [12]. In any case, the presence of water vapor makes useless the utilization of a mask to filter the free radicals [12]. Another example deals with the oxidative functionalization of polypropylene: replacing $O_2$ by water vapor ensures a more homogeneous distribution of the alcohol, carbonyl and carboxyl groups on the surface, but also a functionalization achieved on a larger depth [4].

In biocompatibility applications, the functionalization of hydroxyl groups on polyethylene and polystyrene surfaces has achieved minor success by using either a pure $O_2$ plasma treatment or an Ar plasma treatment followed by an oxygen environment [13, 14]. Water plasma treatment could be considered as an alternative for the preparation of hydroxyl surfaces allowing the covalent immobilization of biologically active molecules and the support of cells colonization [5]. It could also be used for blood compatibility of PTFE prosthesis. For instance, König et al have followed a 3-step process [15] in which a water vapor microwave plasma is used as the first step to facilitate the subsequent post-grafting of acrylic acid on the PTFE surface (2nd step) and the immobilization of the fibrinolytic protein urokinase (3rd step).

In plasma medicine, the role of radicals, such as O, $O_3$, NO and OH, as well as ultraviolet radiation, are still under study for therapeutic applications such as wound healing or dermatology and tumor treatments [6]. For instance, Zhang et al showed the role of O, OH and Ar* radicals on the destruction of cancerous hepatocellular carcimona cells [16]. In this respect, water vapor could be used as a process gas to generate large densities of O and OH radicals.

All of these applications clearly indicate that there are two requirements: Firstly, there is need for the understanding of the chemical mechanisms induced by water vapor at atmospheric pressure and, secondly, for a more specific control of the production of radicals. In this respect, an experimental study is presented in this article where the introduction of water vapor within the post-discharge of an Ar plasma torch is discussed in terms of plasma stability, chemical kinetics and radicals of interest. The results of the present study complement two previous studies published using the same plasma source, namely an experimental investigation by Duluard et al [17] and simulations from Atanasova et al [18].

# Experimental methods

Figure 1 introduces a schematic diagram of the experimental setup, including the plasma source and a water-containing bubbler (at room temperature) dedicated to the injection of water vapor in the post-discharge. The plasma source is an atmospheric RF plasma torch from Surfx Technologies LLC (Atomflo™ 250D), consisting of two parallel circular electrodes perforated by 126 holes, each one with a critical diameter of 0.6 mm. A 126-holes circular metallic mesh is located parallel and downstream to the two electrodes to homogenize the flowing post-discharge. The gas flow is oriented perpendicularly to the two electrodes. As these holes are arranged to form a circular pattern of 20 mm in diameter, this torch is commonly called a 'showerhead plasma torch'. For all of the results discussed in the present article, it operates at a frequency of 27.12 MHz and a power fixed at 80 W. As sketched in figure 1, three regions can be distinguished: (i) the plasma region located between the RF and the ground electrodes, (ii) the inner post-discharge located between the ground electrode and the grid and (iii) the outer post-discharge located beyond the grid.

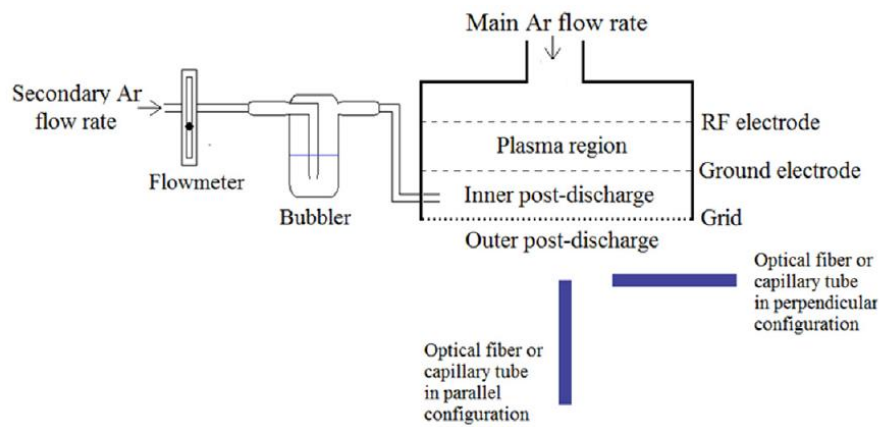

*Figure 1. Schematic diagram of the experimental setup and location of the three discharge regions.*

Two argon inputs were considered: the first one supplies the plasma torch for a flow rate of 30 L min$^{-1}$; it is referenced here as the 'main Ar flow rate'. A second input—assigned here as the 'secondary Ar flow rate', supplies the bubbler via a volumetric flowmeter from Aalborg for flow rates comprised between 0 L min$^{-1}$ and 6 L min$^{-1}$. An Ar/$H_2O$ vapor mixture can therefore be generated and reach the inner post-discharge region via the plasma torch nozzle. Secondary (2nd) Ar flow rates of 0, 2, 4 and 6 L min$^{-1}$ correspond to $H_2O$ flow rates of 0, 0.7, 1.3 and 2 mL s$^{-1}$.

Optical emission spectroscopy (OES) was performed with a Spectra Pro-2500i spectrometer from ACTON research Corporation (0.500 m focal length, triple grating imaging). The light emitted by the post-discharge is collected by an optical fiber set either in a perpendicular configuration or a parallel configuration, as sketched in figure 1. The perpendicular configuration supplies only information on the outer post-discharge, whereas the parallel configuration gives information on the whole post-discharge (inner and outer). The light is then transmitted to the entrance slit (50 µm) of the monochromator where it is collimated, diffracted, focused on the exit slit and finally captured by a CCD camera from Princeton Instruments. Each optical emission spectrum is acquired with a 1800 grooves mm$^{-1}$ grating (blazed at 500 nm) and recorded on 30 accumulations with an exposure time of 25 ms. Experimentally, we have observed that an increase in the $H_2O$ flow rate has always induced a decrease in all the emission lines/bands of the post-discharge. As a consequence, from one water flow rate to another, the de-excitation of a specific species is always proportional to the overall decay of the post-discharge emission. To solve this problem, for every $H_2O$ flow rate, the emissions of all the species were divided by the emission of the whole post-discharge (i.e. a continuum from 250 to 850 nm).

Mass spectra of the post-discharge were acquired with a quadripolar mass spectrometer (Balzers QMS 200) coupled with a turbo molecular pump (Pfeiffer TSU 062H) at a base pressure of approximately $1 \cdot 10^{-8}$ mbar. The mass spectrometer measures the gaseous species in the post-discharge through a capillary tube, set either perpendicular or parallel to the gas flow. The capillary tube is 1 m in length, with an inner diameter of 0.1 mm. It is heated by a resistance to maintain the gas temperature at 150 °C, thereby avoiding condensation before entering the ionization chamber. The ionization energy is set to 70 eV.

DC currents were measured in the post-discharge by placing a copper plate downstream. The distance separating the plasma torch from the copper plate is called the 'gap'. The current is obtained by measuring the voltage difference across a resistor (820 k) connecting the copper plate to an oscilloscope. A digital phosphor oscilloscope from Tektronix (DPO3032) has been used, with a band gap of 300 MHz and a sample rate as high as 2.5 GS s$^{-1}$ for an accurate representation of the signals.

# Results

Figure 2 introduces pictures of the post-discharge operating at an RF power of 80 W, supplied with a main argon flow rate fixed at 30 L min$^{-1}$ and a secondary (2nd) Ar flow rate set to either 2 L min$^{-1}$ or 6 L min$^{-1}$ (with and without water in the bubbler). The amount of $H_2O$ injected in the post-discharge is controlled by the secondary Ar flow rate. If no water is contained in the bubbler, the post-discharge emission is uniform and stable in each hole, whatever the secondary Ar flow rate. However, once water is introduced into the bubbler and carried under vapor via the secondary argon flow rate, the whole emission of the post-discharge vanishes and becomes inhomogeneous through the various holes, randomly changing as a function of time. Moreover, these inhomogeneities become stronger as the amount of water vapor carried by the secondary argon flow rate is increased. Therefore, the inhomogeneities of the post-discharge clearly result from the injected water vapor and not from the secondary Ar flow rate.

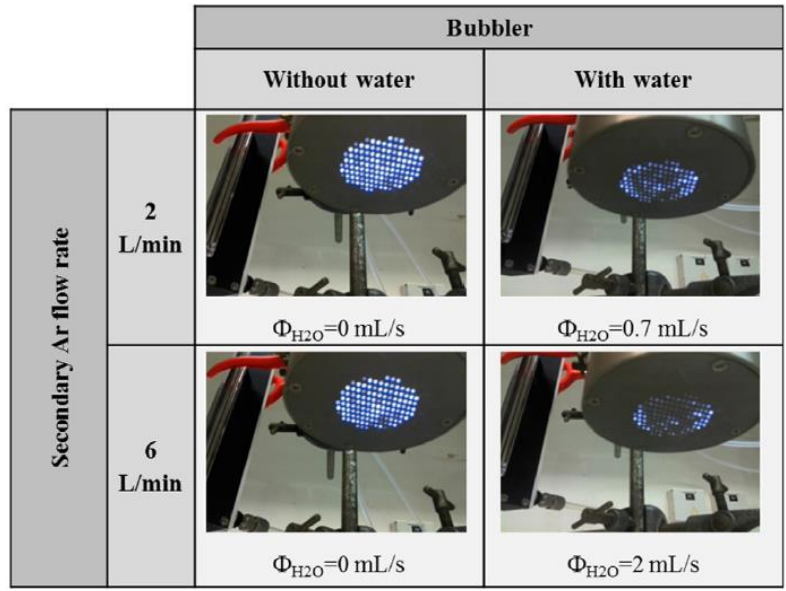

*Figure 2. Influence of the water vapor and secondary Ar flow rates on the apparent post-discharge stability, for P=80 W, Φ(Ar) = 30 L.min$^{-1}$.*

The influence of the $H_2O$ flow rate on the post-discharge's electrical properties was carried out as described in the experimental methods. The post-discharge current as a function of the $H_2O$ flow rate is represented in figure 3. Although some tentative interpretation of this current is given further in this paper (see 'stability' section in the discussion part), and although such current has already been observed and discussed in other papers (but for a torch fed with Helium) [19], these current measurements should only be taken here as qualitative data, hence showing a clear water-depending trend. Its variation shows a strong decrease between 0 and 0.7 mL s$^{-1}$ followed by a slower decay. Without water vapor, the post-discharge is stable, and a current of 32 µA is collected on the copper plate. For $\Phi_{H2O}$ = 2.7 mL s$^{-1}$, a post-discharge current of only 4 µA is measured and corresponds to a lower-emissive and more inhomogeneous post-discharge, as shown in inset. This decrease in the post-discharge current is directly connected to the number of holes ignited in the post-discharge: they are 126 (maximum) without water vapour and less than 50 for $\Phi_{H2O}$ = 2.7 mL s$^{-1}$.

| Species | λ (nm) | Transition | Ref. |
|---|---|---|---|
| OH | 309.1 | $A^2\Sigma^+ \rightarrow X^2\Pi$ | [20, 21] |
| O($^5P$) | 777.4 | $2s^22p^3(^4S^0)3p \rightarrow 2s^22p^3(^4S^0)3s$ | [20, 21] |
| $N_2$ | 337.1 | $C^3\Pi_u \rightarrow B^3\Pi_g$ | [20, 21] |
| Ar ($^3P_2$) | 763.5 | $3s^23p^54p \rightarrow 3s^23p^54s$ | [20, 21] |
| $H_\alpha$ | 656.3 | $3d (^2D) \rightarrow 2p (2P°)$ | [22] |

*Table 1. Species detected by OES.*

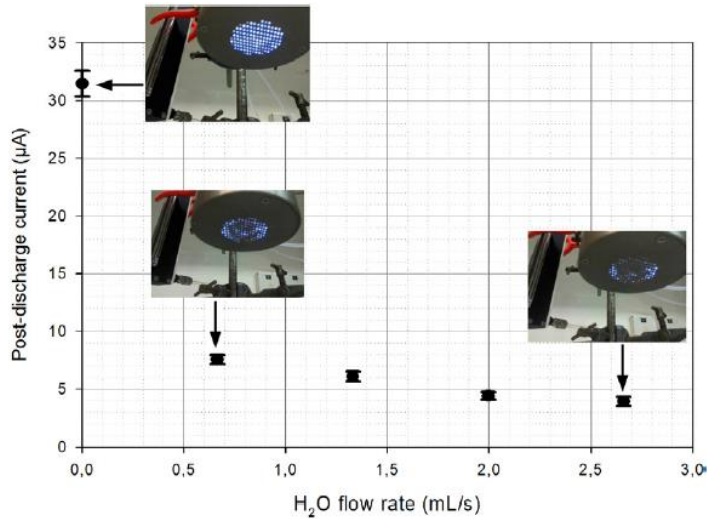

*Figure 3. Influence of the $H_2O$ flow rate on the post-discharge current. Gap = 10 mm. Ar flow rate = 30 L min$^{-1}$, RF power = 80 W.*

The influence of the $H_2O$ flow rate was investigated by measuring the optical emission of several species reported in table 1 namely OH, O, $N_2$ and Ar for the plasma torch supplied with a power of 80 W and an Ar flow rate of 30 L min$^{-1}$. Their variations were plotted as a function of the water flow rate in the perpendicular (figures 4(a) and (c)) and parallel configurations (figures 4(b) and (d)), considering either the influence of the secondary argon flow rate with water vapor (figures 4(a) and (b)) or only the influence of the water vapor (figures 4(c) and (d)) based on the relation $\varepsilon_{dif}\{H_2O\} = \varepsilon\{2^{nd} Ar+H_2O\} - \varepsilon\{2^{nd} Ar\}$ where ε stands for the emission and $\varepsilon_{dif}$ the differential emission. Several observations are noteworthy:

(i) The comparison between figures 4(a) and (b) shows that stronger emissions are obtained in the perpendicular configuration with variations in emission of almost an order of magnitude between 0 and 2.7 mL s$^{-1}$, except for Ar. In this configuration, the production of the OH species seems counterbalanced by the consumption of $N_2$ and O species. Similar trends appear also in the parallel configuration, albeit less significant.

(ii) Figures 4(c) and (d) indicate whether the presence of water vapor in the post-discharge enhances or not the differential emission of the species. If a species differential emission is positive ($\varepsilon_{dif} > 0$), then its density can be considered as more important than in a pure Ar post-discharge (such as OH in the perpendicular configuration). On the contrary, if a species differential emission is negative ($\varepsilon_{dif} < 0$), then its density is higher in the pure Ar post-discharge, such as O, $N_2$ and Ar species in the perpendicular configuration. Figures 4(c) and (d) also indicate whether the injection of water vapor enhances the reactions of production (if $|\varepsilon_{dif}|$ increases versus $\Phi(H_2O)$) or the reactions of consumption (if $|\varepsilon_{dif}|$ decreases with $\Phi(H_2O)$). As an example in the parallel configuration, a rise in $\Phi(H2O)$ increases the differential emission of the Ar line, thus indicating an increase in the density of these excited species.

(iii) The comparison between the {$2^{nd}$ Ar + $H_2O$} and {$H_2O$} cases (i.e. comparison between figures 4(a) and (c) or between (b) and (d)) indicates curves with consistent trends. As an example in the parallel configuration for $\Phi(H_2O)$ = 0–2.7 mL s$^{-1}$, the emission of O decreases from 800 to 250 a.u. in figure 4(a) while from 0 to −330 a.u. in figure 4(c). The reactivity of the emissive species is therefore much stronger in presence of water vapor.

(iv) Fourthly, whatever the graph shown in figure 4, the most significant variations in emission appear when the post-discharge passes from $\Phi(H_2O)$ = 0 to the tiniest injected flow rate. In figure 4(a), the emission of OH increases from 300 to 1350 a.u. for $\Phi(H_2O)$ comprised between 0 and 0.7 mL s$^{-1}$ while in figure 4(c), it increases from 0 to 980 a.u. for $\Phi(H_2O)$ = 0–2.7 mL s$^{-1}$. For higher water flow rates, further variations in intensity are still visible although less significant, as if a plateau was reached (e.g. from 990 to 940 a.u. in the case of OH in figure 4(c)).

(v) A Hα line (656.3 nm) is solely observed in the parallel configuration. Its emission is particularly weak in presence of water vapor, even if it tends to slightly rise by increasing its flow rate.

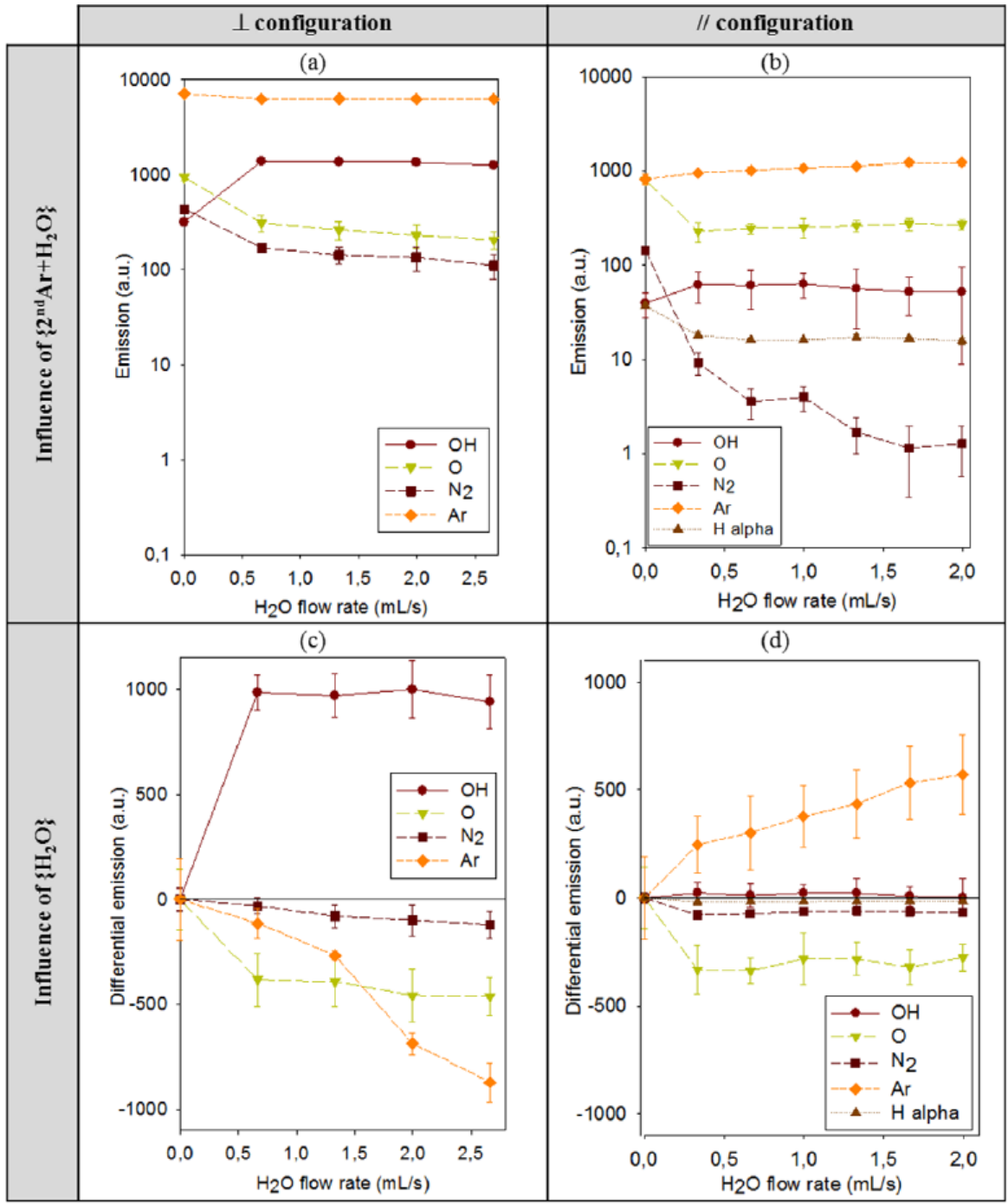

*Figure 4.* OES measurements showing the effect of the $H_2O$ flow rate on the emission and differential emission of OH, O, $N_2$, Ar and H species (Ar flow rate = 30 L min$^{-1}$, power = 80 W). The differential emission is given by $\varepsilon_{dif}\{H_2O\} = \varepsilon\{2^{nd} Ar + H_2O\} - \varepsilon\{2^{nd} Ar\}$

This OES study was completed by taking measurements of the mass spectrometry. The following species were detected where the values in parentheses indicate their corresponding m/z ratios: H (1), $H_2$ (2), N (14), O (16), OH (17), $H_2O$ (18), $N_2$ (28), $O_2$ (32), $H_2O_2$ (34) and Ar (40). The intensity of the species previously detected by OES (OH, O, $N_2$, Ar and H) are plotted in figure 5 and the intensity of the others in figure 6 versus the $H_2O$ flow rate. As in the OES results, each figure comprises four graphs where the perpendicular configuration (a&c), the parallel configuration (b&d), the influence of {$2^{nd}$ Ar + $H_2O$} (a&b) and the influence of {$H_2O$} (c&d) are considered. The aforementioned observations still apply with respect to figure 5: the MS intensities are stronger in the perpendicular configuration, and similar MS intensities trends are obtained between the {$2^{nd}$ Ar + $H_2O$} and {$H_2O$} cases but not necessarily between the ⊥ and // configurations. The most significant variations in emission were detected from 0 to the tiniest measured $\Phi(H_2O)$ flow rate. As in the OES results, the presence of water vapor increases the intensities of OH whilst reducing the intensities of O and $N_2$, whatever the configuration. Furthermore, in the MS results as in the previous OES results, an increase in $\Phi(H_2O)$ induces a lowering of the Ar intensity in the ⊥ configuration while a rising in the // configuration. In figure 6, where species resulting from non-radiative processes are reported, an increase in $\Phi(H_2O)$ indicates a strong increase of both $H_2O$ and $O_2$, a slight increase of both $H_2O_2$ and $H_2$ and, finally, a decrease in N, whatever the configuration.

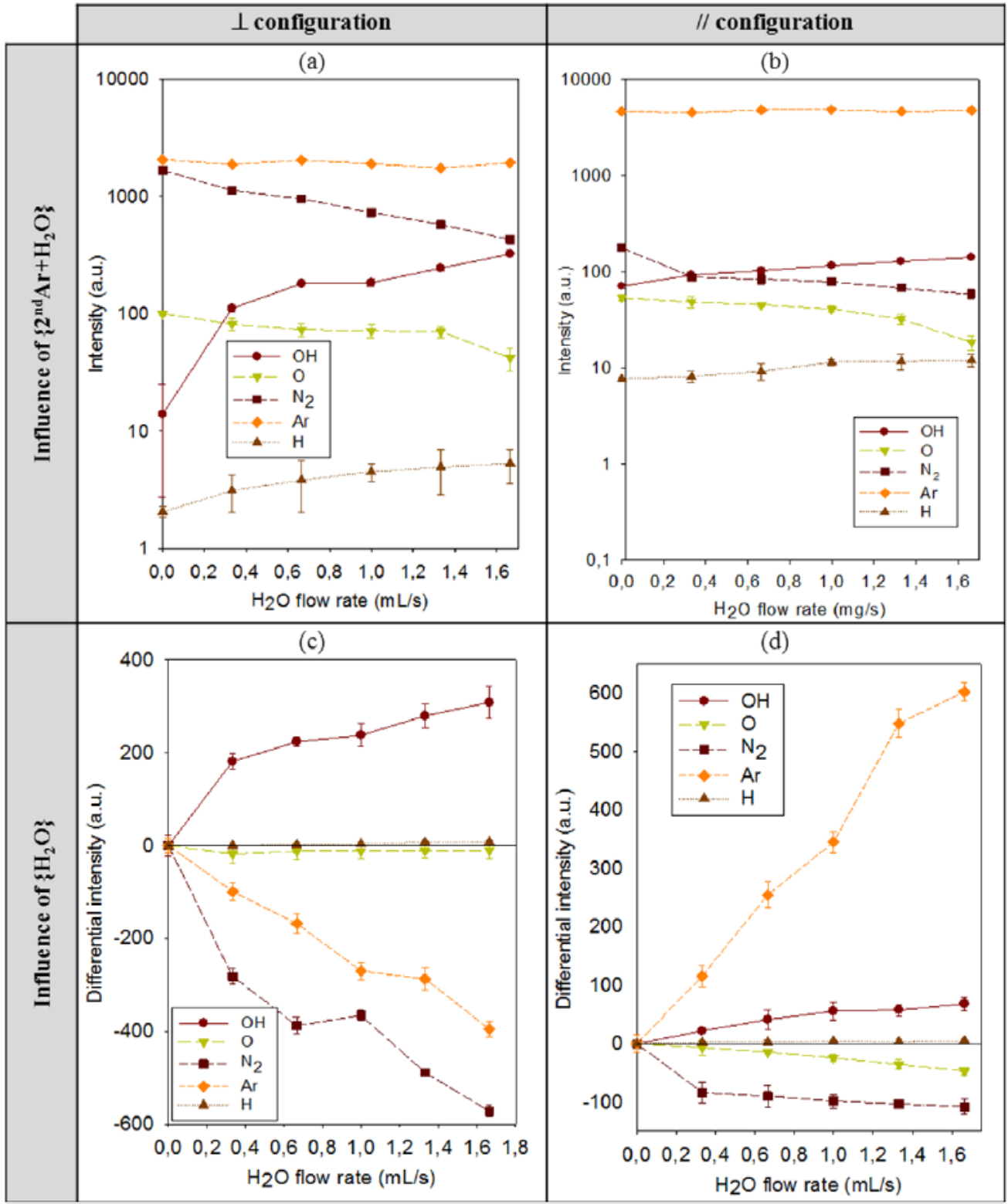

Figure 5. Mass spectrometry measurements showing the effect of the $H_2O$ flow rate on the intensity and differential intensity of OH, O, $N_2$, Ar and H species (Ar flow rate = 30 L min$^{-1}$, power = 80 W).

# Discussion

The discussion is focused on the possible reactions occurring in the interelectrode and post-discharge regions for each species detected by OES and/or mass spectrometry. The stability of the post-discharge depending on the water vapor injected is also discussed.

## Argon species in the interelectrode region

In the interelectrode region, the production of non-ionized excited Ar species is expected through elementary processes such as elastic scattering [R-1], excitation [R-2] and recombination [R-3] (all the R-X reactions are described in table 2). The production of Ar+ ions is achieved through the direct ionization reaction [R-4], as well as the stepwise ionization, described by reaction [R-5] and the Penning ionization resulting from the collision of two excited Ar atoms [R-6]. Besides this, the production of Ar+2 ions is also expected in the interelectrode region through the reactions [R-7], as supported by the simulations of Atanasova et al [18]. The production of Ar metastable species has also been evidenced through either electron collision [R-8] and Ar+ collision [R-9] with Ar [23].

| | # | Reaction | Rate constant |
|---|---|---|---|
| Inter-electrode region | [R-1] | $e + Ar \rightarrow e + Ar$ | a |
| | [R-2] | $e + Ar \rightarrow e + Ar^*$ | $k = 3.71 \cdot 10^{-8} \cdot \exp\left(-\frac{15.06}{T_e}\right)$ |
| | [R-3] | $Ar^+ + Ar + e \rightarrow Ar^* + Ar$ | $k = 1.01 \cdot 10^{-11}$ |
| | [R-4] | $Ar + e \rightarrow Ar^+ + 2e$ | $k = 1.23 \cdot 10^{-7} \cdot \exp\left(-\frac{18.68}{T_e}\right)$ |
| | [R-5] | $Ar^* + e \rightarrow Ar^+ + 2e$ | $k = 2.05 \cdot 10^{-7} \cdot \exp\left(-\frac{4.95}{T_e}\right)$ |
| | [R-6] | $Ar^* + Ar^* \rightarrow Ar^+ + e + Ar$ | $k = 6.2 \cdot 10^{-10}$ |
| | [R-7] | $2Ar + Ar^+ \rightarrow Ar + Ar_2^+$ | $k = 1.7 \cdot 10^{-32}$ |
| | [R-8] | $Ar + e \rightarrow Ar^M + e$ | a |
| | [R-9] | $Ar + Ar^+ \rightarrow Ar^M + Ar^*$ | a |
| Post-discharge region | [R-10] | $Ar^M + Ar \rightarrow Ar^* + Ar^*$ | $k = 2.3 \cdot 10^{-21}$ |
| | [R-11] | $Ar^M + H_2O \rightarrow Ar + OH + H$ | $k = 4.5 \cdot 10^{-10}$ |
| | [R-12] | $O(^1D) + H_2O \rightarrow OH + OH$ | $k = 2.2 \cdot 10^{-10}$ |
| | [R-13] | $O(^3P) + H_2O \rightarrow OH + OH$ | $k = 2.5 \cdot 10^{-14}$ |
| | [R-14] | $O + H + N_2 \rightarrow OH + N_2$ | $k = 1.0 \cdot 10^{-27}$ |
| | [R-15] | $OH + OH + N_2 \rightarrow H_2O_2 + N_2$ | $k = 1.6 \cdot 10^{-23}$ |
| | [R-16] | $H_2O + N_2^* \rightarrow H_2O + N_2$ | $k = 3.9 \cdot 10^{-10}$ |
| | [R-17] | $O(^3P) + OH \rightarrow O_2 + H$ | $k = 2.3 \cdot 10^{-11}$ |
| | [R-18] | $H_2O + O_3 \rightarrow H_2O_2 + O_2$ | $k = 1.1 \cdot 10^{-22}$ |
| | [R-19] | $H + O_3 \rightarrow OH + O_2$ | $k = 1.4 \cdot 10^{-10}$ |
| | [R-20] | $O_2 \rightarrow O + O$ | $k = 1.01 \cdot 10^{-8}$ |
| | [R-21] | $OH + OH \rightarrow H_2O + O$ | $k = 2.5 \cdot 10^{-15}$ |
| | [R-22] | $H_2O + O \rightarrow H_2O_2$ | a |

Table 2. Plausible reactions occurring in the Ar/H$_2$O post-discharge (units: cm$^3$.s$^{-1}$ for two-body collisions and cm$^6$.s$^{-1}$ for three-body collisions).

## Argon species in the post-discharge

The plasma torch is supplied in argon via two inputs: the first one allows striking and sustaining the discharge in the interelectrode region, while the second plays as a carrier gas for the water vapor in the inner post-discharge region. As shown in figures 4(a) and (b), a rise in the water vapor flow rate does not lead to a significant variation in the Ar emission, i.e. the excitation of Ar. However, the sole contribution of the water vapor, as highlighted in figures 4(c) and (d), shows a discrepancy in the emission of Ar: a decrease in the perpendicular configuration (outer region) with an increase in the parallel configuration (inner region). Similar behavior has been obtained with mass spectrometry, as shown in figures 5(c) and (d). Two possibilities may explain these patterns:

(i) the consumption of the Ar metastable species by the water vapor, as suggested by the reaction [R-11]. An increase in the $H_2O$ reactant may consume more efficiently the Ar metastable species, thus resulting in one hand in the dissociation of $H_2O$ into OH and H radicals as evidenced in the figure 4 and, on the other hand, in the excitation of Ar. As $\Phi(H_2O)$ increases, the emission of excited Ar species decays in the outer post-discharge due to the upwards consumption of the ArM which can no longer transfer part of their energy via [R-10]. As a result, the more water vapor is injected, the more the Ar species can be excited and confined in the inner post-discharge region.

(ii) injecting water vapor in the inner post-discharge may change the fluid dynamics of the main argon flow since it is injected perpendicularly. The resulting spatial distribution of the flow through the 126 holes grid would become more inhomogeneous by increasing the water vapor, i.e. becoming stronger through the holes located in the grid's central region and lower through the holes in periphery.

Simulations have been conducted by Atanasova et al on the current plasma torch for an argon flow rate set at 30 L min$^{-1}$ [18]. They have revealed the existence of $Ar_2^+$ ions in the post-discharge, especially through dissociative recombination (see [R-7]) while no $Ar^+$ ion has been calculated. For instance, 2 mm away from the ground electrode, the density of $Ar^+$ is negligible while it is $5 \cdot 10^{16}$ m$^{-3}$ for $Ar_2^+$. As expected, $Ar^+$ ions has not been detected by OES. The $Ar_2^+$ ions have not been observed either due to the limited range of our optical spectrometer. Therefore, even if these ions could not be evidenced experimentally for material reasons, their existence in the first millimeter of the post-discharge is at least sustained by the simulations of Atanasova.

Two last observations on the OES and MS results are noteworthy:

(i) On the whole $\Phi(H_2O)$ range, the intensities of Ar are higher in figure 4(a) (approximately 8000 a.u.) than in figure 4(b) (approximately 1000 a.u.). The reason for this is that even if the optical fiber presents the same solid angle of collection independently of the configuration, the integration of the light depends on the configuration. As illustrated in figure 7, light is collected over 40 holes in the perpendicular configuration whereas approximately 8 holes in the parallel configuration. This explains also why the uncertainties on the data points are stronger in figure 4(b) than in figure 4(a).

(ii) The Ar mass spectra from figure 6 present similar trends of those obtained by OES. The intensities measured are however higher in the parallel configuration (see figure 6(b)) since the issue is no more dealing with a solid angle of collection but with the section of the capillary tube which collects more species when frontally exposed to the post-discharge flow.

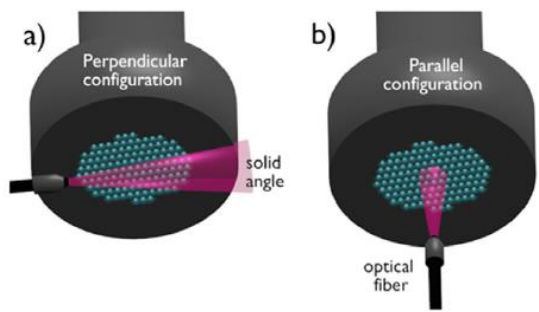

*Figure 7. Regions analyzed in (a) the perpendicular configuration and (b) the parallel configuration.*

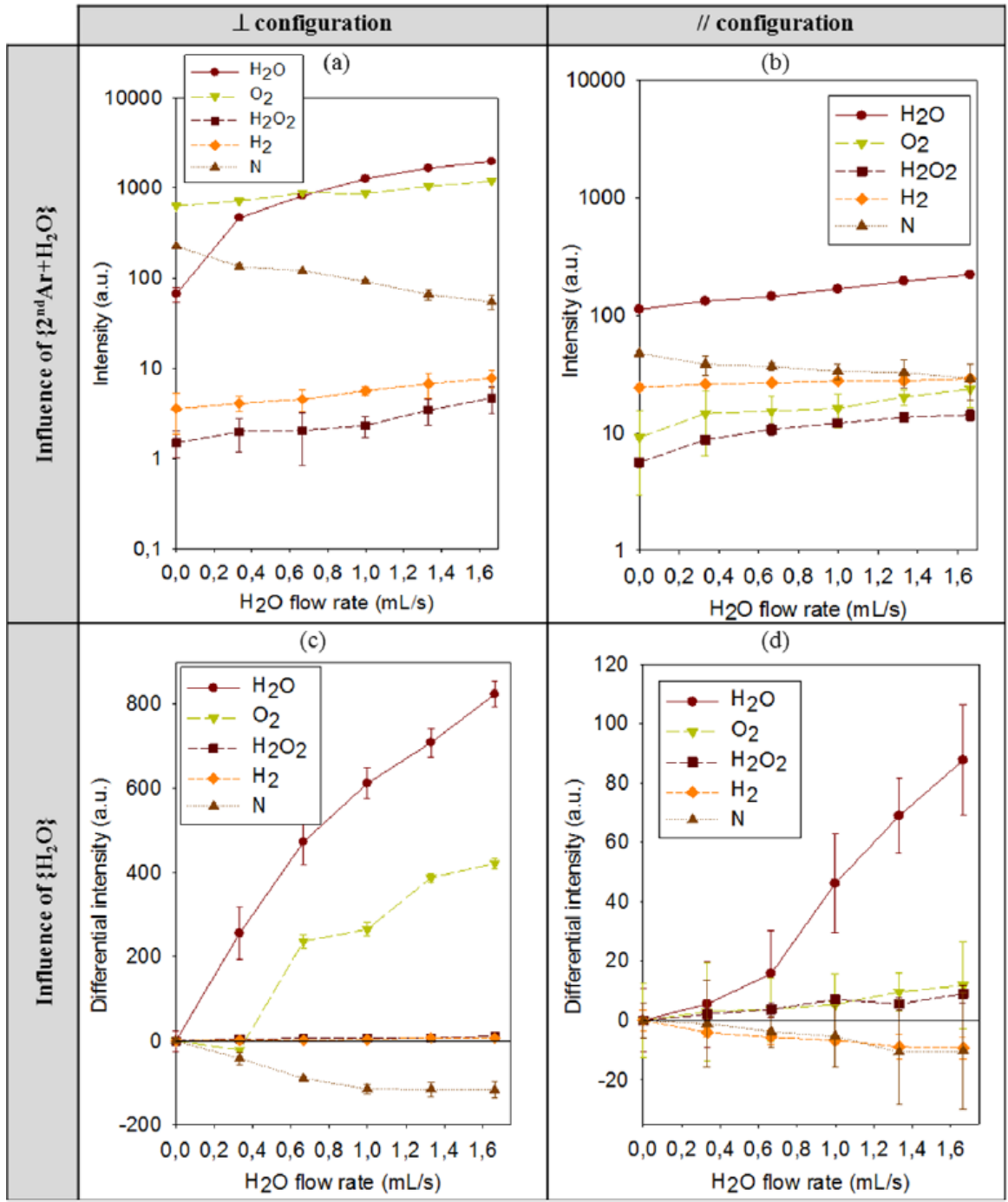

*Figure 6. Mass spectrometry measurements showing the effect of the $H_2O$ flow rate on the intensity and differential intensity of $H_2O$, $O_2$, $H_2O_2$, $H_2$ and N (Ar flow rate of 30 L min$^{-1}$ and RF power of 80 W).*

## Electrons

The mean free path of an electron in the ambient air with an energy comprised between 20 and 100 eV is approximately 0.5 µm [24]. As in atmospheric post-discharges the electron temperature is usually lower than 2 eV [25], the corresponding mean free path is even much smaller than 0.5 µm. Owing to their very short mean free paths at atmospheric pressure, and to the thickness of the ground electrode (at least 1 mm), the presence of free electrons in the inner post-discharge region seems a priori highly hypothetical. However, the simulations of Atanasova et al conducted on this plasma torch indicate that, due to the effect of the flowing carrier gas, an electron density as low as $5 \cdot 10^{10}$ cm$^{-3}$ can be obtained 1 mm away from the ground electrode before it drastically drops to zero before reaching the grid [18]. Therefore, a very low electron density may be expected in the inner region of the post-discharge even if its role seems negligible for the excitation of Ar or the dissociation of $H_2O$ owing to the too low energy of the electrons.

## $H_2O$ molecules

Water vapor cannot be detected in the interelectrode region because of the configuration of the plasma torch. In the flowing post-discharge, two sources of water can be distinguished: the water vapor injected via the bubbler (source controlled) and the water vapor from the ambient air (source uncontrolled). The optical emission of $H_2O$ molecules cannot be detected since occurring at 6260 nm [33] and 11876 nm [34]. However, mass spectrometry measurements revealed an increase in the intensity of $H_2O$ with $\Phi(H_2O)$. As this increase is higher than when the plasma torch is off, it thus indicates a partial dissociation of the water vapor through at least one of the following reactions leading to the production of OH radicals, i.e. [R-11], [R-12] and [R-13].

## Nitrogen molecules

Both OES and MS measurements have evidenced the presence of molecular nitrogen in the post-discharge, due to the ambient air. In the perpendicular configuration, a gradual drop in the $N_2$ optical emission and MS peak intensity was observed as a function of the water vapor flow rate (see figures 4(c) and 5(c)). Molecular nitrogen can be deexcited by the dissociation products of $H_2O$, i.e. OH and O, as previously stated through reactions [R-14] and [R-15]. Besides, MS measurements have evidenced a partial consumption of water in the discharge, meaning that $N^*_2$ might also be quenched by $H_2O$ molecules ([R-16]). Reactions leading to the dissociation of $N_2$ or to the formation of $N^+$ ions are irrelevant since neither N nor N+ emissions could be observed in the optical range. Furthermore, no optical emission of $N_2^+$ (15.58 eV) could be observed since Penning ionization of $N_2$ cannot be induced by the too-low energetic $Ar^M$ species (11.55 eV) in the post-discharge [35–37].

In the parallel configuration, the drop of the $N_2$ emission (figure 5(c)) or $N_2$ MS intensity (figure 5(d)) only occurs when a tiny flow of water is injected (0.33 mL s$^{-1}$). In our setup, a further increase in the water vapor flow rate remains too low to detect, by OES or MS, significant variations of $N_2$.

## OH radicals

For $\Phi(H_2O)$ = 0 ml s$^{-1}$, the production of these OH radicals can result from the interaction of the post-discharge with the ambient atmosphere, where water vapor can be dissociated as shown by reaction [R-12], involving an excited oxygen radical as collisional partner. The production of OH radicals through electron impact processes in the post-discharge is highly hypothetical due to the too low energy and density of the electrons [38–40]. Moreover, the production of OH radicals in the interelectrode region is not expected since the plasma torch is supplied in argon and eventually in oxygen: no H source is present.

For $\Phi(H_2O) > 0$ mL s$^{-1}$, the production of OH can be understood combining the trends of OH and O emissions in figure 4(c) (perpendicular configuration). From 0 to 0.7 mL s$^{-1}$, the consumption of atomic O is counterbalanced by the production of OH radicals, whereas beyond 0.7 mL s$^{-1}$ an equilibrium seems to be reached for the consumption of O radicals; hence, there is a limitation in the emission of the OH radicals (around 1000 a.u.). As a result, the previous reaction [R-12] still occurs. Also, MS results in figure 5(c) indicate that at least one non-radiative process may also be responsible for their production since the intensity of OH increases as a function of the water vapor flow rate (while the emission of OH remains constant in figure 4(c)). As an example of this, the reaction [R-13] involving ground state oxygen atoms could be responsible for the production of non-emitting OH species.

## H radicals

The comparison between figures 4(c) and (d) shows that the production of H radicals through radiative processes has only been evidenced in parallel configuration; the mass spectrometry measurements indicate the production of H radicals both in ⊥ and // configurations (see figures 6(c) and (d)).

In the post-discharge, the injection of water vapor does not seem to favor the production of H radicals since no significant variation has been observed in the Hα emission in figure 4(d) and in its MS intensity in figure 5(d). However, variations can be observed in figures 4(b) and 5(b) where the influence of the secondary argon flow is taken into account, meaning that new pathways such as [R-11] and [R-17] may barely happen in the post-discharge. It is also worth mentioning that OES has not revealed any formation of molecular hydrogen since no Fulcher band could be detected between 590 nm and 640 nm irrespective of the setup configuration [41, 42]. The molecular hydrogen detected by MS (see figure 5) may result from the recombination of H atoms in the ionization chamber.

## O radicals and $O_2$ molecules

The production of $O_2^+$ ions has already been observed in the case of flowing post-discharges supplied in helium as a carrier gas [19]. These ions were the result of Penning ionization induced by He metastable species. Here, the Penning ionization of $O_2$ molecules by Ar metastable species can barely be suggested since no production of $O_2^+$ ions could be evidenced by OES. The reason for this is that the energetic level of $Ar^M$ species is lower than the energy required to ionize $O_2^+$ ions. For the same reason, no $O_2$ metastable species could be evidenced since no band of the singlet sigma metastable oxygen $O_2(b^1\Sigma_g^+)$ has been detected between 758 and 770 nm [19]. However, the presence of molecular oxygen has been highlighted by mass spectrometry as increasing with $\Phi(H_2O)$ irrespective of the configuration (see figure 6). Even without water injected, the figures 6(a) and (b) clearly indicate a peak intensity of $O_2$ specific to molecular oxygen from the ambient atmosphere. An increase in $\Phi(H_2O)$ is responsible for the production of $O_2$ through non-radiative processes such as [R-17], [R-18] and [R-19]. The ozone species required by the two latter reactions have already been measured in the outer post-discharge at concentrations comprised between $10^{14}$ and $10^{15}$ cm$^{-3}$ by investigating the absorption of the mercury emission line [17].

One of the main pathways explaining the efficient production of atomic oxygen in the post-discharge is the direct dissociation of molecular oxygen contained in the ambient atmosphere (see [R-20]). However, another reaction—even if less obvious—should also be taken into account. Indeed, we have already stated that atomic oxygen could control the production of OH radicals essentially through reactions [R-12] and [R-13]. We have also noticed in figure 4(c) that any further increase in water vapor flow rate induced limitations of OH and O emissions. Those two plateaus may evidence an equilibrium with respect to the production/consumption of O and OH species, resulting from a reaction opposite to [R-12] and [R-13], namely [R-21]. In this reaction, two OH radicals could lead to the production of water and O radicals.

### $H_2O_2$

The formation of hydrogen peroxide was detected by mass spectrometry in the ⊥ and // configurations. Without injecting water vapor, a very tiny amount of $H_2O_2$ was detected and it may have resulted from [R-15] where two OH radicals in presence of molecular nitrogen leads to the production of $H_2O_2$. By injecting water vapor, two other reactions may enhance the mass spectrum intensity of $H_2O_2$ as shown in figures 6(a) and (b), namely reactions [R-18] and [R-22], with the first implying ozone molecules and the second one oxygen radicals.

### Stability

As shown in figure 1, the post-discharge stability is weakened as the water vapor flow rate is increased. The main reaction that could explain this trend is [R-11] where Ar metastable species are consumed by water. This consumption of Ar metastables can also be indirectly evidenced through the decrease in the post-discharge current versus the water vapor flow rate (see figure 3). Indeed, Bogaerts et al have shown that for a DC argon glow discharge, $Ar_2^+$ ions are predominantly formed by associative ionization, both through highly excited Ar** atoms (Hornback-Molar process) and by metastable-metastable collisions [43]. As [R-11] consumes the Ar metastable atoms, the $Ar_2^+$ ions can no longer be formed through this reaction, and so there is a partial decrease of the current intensity with $\Phi(H_2O)$.

# Conclusion

The reactivity of water vapor introduced in the flowing post-discharge an atmospheric plasma torch was investigated firstly by correlating electrical characterization, OES and MS results and previous studies achieved with the same plasma source (one dealing with simulations and the other with experiments). Inhomogeneities were explained as resulting from the consumption of the Ar metastable species by $H_2O$ molecules. The gradual consumption in $Ar_2^+$ ions was advanced as an indirect consequence of this consumption in Ar metastables. Moreover, we have also shown the absence of any Penning ionization of $O_2$ and $N_2$ observed in a flowing helium post-discharge. The production and consumption reactions of O, OH and H radicals have been shown through various reactions, in particular the interdependency between O and OH. Finally, the interaction of the post-discharge with the ambient air has also been evidenced since molecular nitrogen was detected by OES.

# Acknowledgment

This work was part of the Interuniversity Attraction Pole (I.A.P) programs 'PSI _ Physical Chemistry of Plasma Surface Interactions—IAP-VII/12, P7/34—financially supported by the Belgian Federal Office for Science Policy (BELSPO).